# PEARL: A Task-Aware Framework for Evaluating Differentially Private Synthetic Educational Data

**Xianghui MENG, Yujing ZHANG, Jionghao LIN***
*The University of Hong Kong, Hong Kong SAR, China*
*jionghao@hku.hk

**Abstract:** Personalized learning systems rely on real learner data, including performance, behavior, and demographic information, but these data are highly privacy-sensitive. Differentially private (DP) synthetic data can support system development and educational research while reducing exposure of individual learners. Existing evaluations, however, assess privacy and predictive usefulness separately, without determining whether synthetic learner data remain usable for the intended personalized learning task. We introduce PEARL (Privacy-Equivalence Audit and Release Ledger), which approves a DP synthetic educational dataset only when it passes all required checks of validity, privacy protection, predictive usefulness, and suitability for the intended educational task, while recording why each rejected dataset fails. Across 96 study settings, each defined by a dataset, data-generation method, privacy budget, and random seed, only 12 produced synthetic datasets that passed all applicable PEARL checks. Many privacy-protected datasets were rejected for omitting important outcome groups, such as withdrawn students, or for failing to preserve the order of learning activities. Fairness analysis further showed that some datasets passing privacy and predictive-usefulness checks still yielded unequal at-risk prediction performance across groups defined by disability and socioeconomic background. Moreover, Deep Knowledge Tracing and Self-Attentive Knowledge Tracing learned no meaningful next-response patterns from any tested synthetic knowledge-tracing dataset, showing that privacy protection alone does not guarantee usefulness for dropout prediction, knowledge tracing, or adaptive tutoring.

## 1. INTRODUCTION

Personalized learning systems rely on real learner data to understand individual needs and provide personalized support. These data may include information about learners' performance, behaviors, and demographic backgrounds, which can be used to develop and evaluate applications such as dropout prediction, knowledge tracing, and adaptive tutoring. Sharing and reusing learner data can therefore help researchers and system developers identify learning patterns, evaluate educational models, and improve personalized learning technologies. However, learner data are highly privacy-sensitive because they may reveal students' identities and other personal characteristics. Educational institutions must therefore balance the benefits of making learner data available for research and system development with their responsibility to protect individual learners.

Differentially private (DP) synthetic data offer a potential way to address this tension. Instead of sharing original learner records, institutions can generate artificial records that reproduce relevant patterns in the original data. Differential privacy further limits how much any individual learner's record can influence the generated dataset, thereby reducing the risk that information about a particular learner can be recovered from the shared data (Dwork & Roth, 2014). In principle, differential privacy can support educational research and system development while limiting the direct exposure of original learner records (Liu et al., 2025). However, the central question is whether the resulting synthetic data both protect learner privacy and retain the information required for their intended educational use.

Existing evaluations of DP synthetic data often report privacy protection and predictive usefulness as separate outcomes (Stadler et al., 2022; Tao et al., 2021). Reporting these outcomes separately does not determine whether a synthetic dataset is suitable for a particular personalized learning task. This limitation matters for educational data because usefulness depends on more than overall predictive performance. Student datasets often contain uncommon outcome groups, such as early withdrawal; a generation method may remove these groups and leave the data unsuitable for identifying at-risk students. Clickstream data contain ordered learning activities; a synthetic dataset may preserve activity frequencies while losing temporal order, preventing sequence models from learning how behavior changes over time. Synthetic data may also preserve overall predictive performance while producing less accurate at-risk predictions for particular demographic or socioeconomic groups. These problems may not be detected when privacy and predictive usefulness are evaluated only through separate aggregate scores.

To address this gap, we introduce PEARL (Privacy-Equivalence Audit and Release Ledger), a task-aware evaluation framework for differentially private synthetic educational data. PEARL approves a synthetic dataset only when it passes all checks applicable to its intended educational use: validity of the privacy assessment, protection of individual learners, predictive usefulness with retention of important outcome groups, and support for the intended educational task. For datasets containing protected attributes, PEARL further examines whether at-risk prediction performance differs substantially across student groups. Rather than presenting separate scores for ad hoc judgment, PEARL produces an explicit approval or rejection under predefined criteria and maintains a release ledger recording the failed check and reason for each rejection. We evaluate PEARL on six public educational datasets representing tabular student records, MOOC clickstream sequences, and knowledge-tracing response sequences (Section 3.1). The evaluation covers 96 study settings, each defined by a dataset, data-generation method, privacy budget, and random seed. Only 12 settings produced synthetic datasets that passed all applicable checks. This evaluation identifies which settings yield approved releases and why other privacy-protected datasets remain unsuitable for their intended uses. The study addresses the following Research Questions (**RQ**s):

- **RQ1**: Does the PEARL validity check confirm that the membership inference evaluation avoids producing misleading privacy signals?
- **RQ2**: Which synthetic educational releases pass the full PEARL audit, and where do rejected releases fail?
- **RQ3**: When synthetic datasets pass privacy and standard predictive-usefulness checks, do they still support their intended educational tasks, particularly clickstream modeling and knowledge tracing?

This paper makes three contributions. *First*, it introduces PEARL as a task-aware framework that integrates validity, privacy protection, predictive usefulness, and task-specific suitability into a transparent approval or rejection decision, with a release ledger and additional analyses of subgroup fairness and stronger membership inference attacks. *Second*, a systematic evaluation of 96 settings across six educational datasets shows that many privacy-protected synthetic datasets still omit important outcome groups, distort at-risk prediction across student groups, produce unreliable outcome patterns, or lose essential sequence information. *Third*, it shows that usefulness depends on the intended task: MST and DP-CTGAN produce approved tabular releases in some settings, and DP-Markov produces approved ACT-MOOC clickstream releases, whereas none of the tested synthetic knowledge-tracing datasets preserve sufficient next-response patterns for Deep Knowledge Tracing (DKT) or Self-Attentive Knowledge Tracing (SAKT). Privacy protection alone is therefore insufficient evidence that synthetic data can support dropout prediction, knowledge tracing, or adaptive tutoring.

## 2. RELATED WORK

### *2.1 Differentially Private Synthetic Data Generation*

Differential privacy provides a formal guarantee that limits how much any individual record can change the output of a data release mechanism (Dwork & Roth, 2014). When applied to synthetic data generation, it requires controlled randomness so that including or removing one learner has only a limited effect on the resulting synthetic dataset. Stronger privacy generally requires more noise and can reduce fidelity to the original patterns. Existing DP synthesizers fall into two broad families. Count-based methods, such as DP-Histogram, DP-Marginals, and MST, perturb frequency tables or low-dimensional summary statistics (Bowen & Liu, 2020; Tao et al., 2021). Neural methods, including DP-CTGAN, capture more complex distributions by training a generative adversarial network with gradient clipping and Gaussian noise (Abadi et al., 2016; Fang et al., 2022; Xu et al., 2019). Recent educational work has begun to examine private synthesis for sharing real-world learner data, including cyclic adaptive private synthesis for small, high-dimensional educational datasets (Ito et al., 2026). Most broader evaluations still focus on census or health data, leaving performance on educational data less understood.

### *2.2 Privacy Challenges in Educational and Sequential Data*

Educational data pose challenges that general synthetic-data evaluations may overlook. Student records can include important outcome groups such as dropout (Realinho et al., 2021), and privacy-preserving generation may fail to retain these groups. Sequential learning logs add a further challenge: because each student contributes an ordered history of events, privacy risk should be measured at the student level rather than the event level (Chen et al., 2012; Kellaris et al., 2014; Miranda-Pascual et al., 2023). Knowledge-tracing data are especially demanding because DKT (Piech et al., 2015) and SAKT (Pandey & Karypis, 2019) learn from ordered skill-response sequences. Recent work also studies user-level differential privacy directly in knowledge-tracing models (Kabir et al., 2025). Synthetic data must therefore preserve this sequence structure, which existing privacy and predictive-usefulness evaluations do not directly assess. PEARL addresses this gap.

### *2.3 Membership Inference Attacks on Synthetic Data*

A membership inference attack (MIA) tests whether a specific record was part of the data used to produce a released dataset or model (Yeom et al., 2018). For synthetic data, the standard screening approach is nearest-neighbor MIA: a query record receives a higher membership score when it is unusually close to a synthetic record (Chen et al., 2020; Stadler et al., 2022; van Breugel et al., 2023). Attack performance is reported as AUC, where 0.5 indicates chance-level discrimination (Fawcett, 2006). We use nearest-neighbor MIA as a reproducible screening test, not proof against all attacks; stronger attacks have been developed for machine-learning models generally and for synthetic-data releases specifically (Liu et al., 2023; Ward et al., 2025).

### *2.4 Privacy-Utility Evaluation and the Missing Educational Benchmark*

Prior privacy–utility benchmarks combine MIA results with downstream prediction scores and fidelity metrics such as TSTR (train-on-synthetic-test-on-real) retention (Davila Restrepo et al., 2025; Du & Li, 2025; Stadler et al., 2022; Tao et al., 2021). Recent educational synthetic-data releases likewise report privacy, fidelity, and analytical utility as parallel validation dimensions (Agal, 2026). Reported side by side, these metrics do not answer whether a specific release should be certified: a release may pass a privacy screen while losing target classes, inflating utility through label artifacts, or failing sequence-based models.

## 3. METHOD

### *3.1 Datasets*

We evaluate PEARL on six public educational datasets covering tabular student records, sequential MOOC clickstreams, and knowledge-tracing response sequences. The primary evaluation includes four datasets. (i) UCI 697, Predict Students' Dropout and Academic Success (Realinho et al., 2021), has three outcome classes; after preprocessing, our evaluation uses 2,100 higher-education student records. (ii) UCI 856, Higher Education Students Performance Evaluation (Yilmaz & Şekeroğlu, 2019), is used as a multi-class low-data stress case; after preprocessing, our evaluation uses 101 records. (iii) OULAD (Kuzilek et al., 2017; Knowledge Media Institute, 2015) contributes 1,820 students across 2,121 enrollment rows after preprocessing, with four outcome categories; records are split at the student level because one student may appear in multiple terms. (iv) ACT-MOOC (Kumar et al., 2019) is a temporal MOOC interaction dataset with dropout information; the processed evaluation set contains 2,000 learners and 121,682 clickstream events, and membership inference is evaluated at the learner level. ASSISTments 2009-2010 Skill Builder (Feng et al., 2009) and EdNet KT1 (Choi et al., 2020) serve as knowledge-tracing extensions. After sequence preprocessing, they contain 346,860 and 539,460 events, respectively. Sequences are capped at 200 events, the item vocabulary is capped at 512, and fixed-seed train-validation-test splits are used.

### *3.2 Synthesizers and Privacy Mechanisms*

For tabular data, Gaussian Copula and CTGAN serve as non-private baselines. We evaluate four DP methods: DP-Histogram and DP-Marginals (noisy frequency or low-dimensional statistics), MST (selected and perturbed variable relationships), and DP-CTGAN (conditional GAN with gradient clipping and calibrated Gaussian noise). PATE-GAN (Jordon et al., 2019) is an additional DP neural baseline in the subgroup-fairness and stronger-attack analyses (Section 4). Tabular DP methods use $\varepsilon \in \{0.5, 1.0, 2.0, 5.0, 10.0\}$. For sequential data, DP-Markov generates sequences from noisy first-order transitions over clipped user histories (Wang et al., 2022), while DP-HistSeq synthesizes order-free event bags. Sequential methods use $\varepsilon \in \{1.0, 5.0, 10.0\}$.

### *3.3 PEARL Evaluation and Approval Protocol*

Figure 1 summarizes the PEARL evaluation and approval protocol. Each synthetic dataset is assessed through four sequential checks: **Validity**, which confirms that randomized membership labels do not produce misleading membership-inference results; **Privacy**, which screens for membership disclosure and exact-copy leakage; **Utility**, which examines whether the data retain target groups and support downstream prediction; and **Task-Specific Suitability**, which verifies whether the data support the intended educational use. For tabular datasets containing protected attributes, PEARL applies a subgroup-fairness check. The release ledger records the outcome of each check and the reason for every rejection.

A dataset is approved only when it passes every check applicable to its data type. The validity check examines whether membership inference produces misleading results when no true membership information is present. We randomly shuffle membership labels and require the mean attack AUC to remain within 0.10 of chance: $|AUC - 0.5| < 0.10$. Above-chance scores under randomized labels would indicate that the evaluation procedure is unreliable for release decisions. This gate is applied to OULAD, UCI 697, UCI 856, and ACT-MOOC at the appropriate row or learner level. The privacy check evaluates whether an attacker can determine whether a learner's data were used to generate the synthetic dataset. For tabular datasets, PEARL runs row-level nearest-neighbor MIA; for sequential data such as ACT-MOOC, it runs user-level MIA over learner sequence representations. A release fails if MIA AUC exceeds 0.55 or if any records are exact copies of the original data. The 0.55 threshold is a benchmark screening rule, not a universal privacy guarantee.

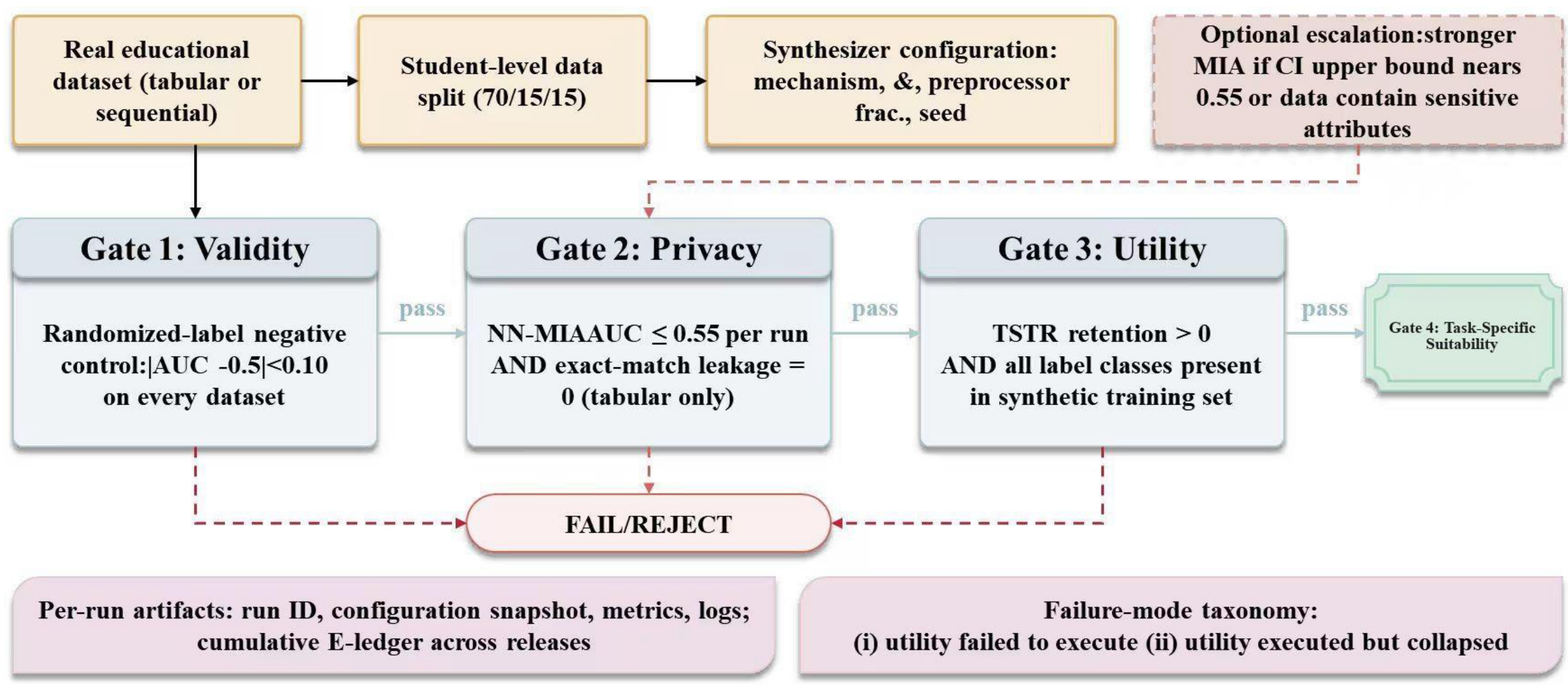

*Figure 1.* The PEARL evaluation and approval process.

The utility check examines whether a privacy-passing synthetic dataset can support downstream prediction. Models are trained on synthetic data and tested on held-out real data, and their macro-F1 is compared with models trained on real data. Evaluated classifiers include logistic regression, random forest, multilayer perceptron, and XGBoost. A synthetic dataset is rejected if it omits any target outcome group, produces no predictive performance, or cannot be compared with a usable real-data baseline. For sequential and knowledge-tracing settings, PEARL applies task-specific suitability as a fourth check. ACT-MOOC uses a GRUSeq retention check to test whether temporal order is preserved for sequence modeling. ASSISTments and EdNet KT1 are evaluated with DKT and SAKT for next-step correctness prediction; they are task-specific extensions because not all certification gates apply. The release ledger records each rejection with its gate identifier and failure reason.

### *3.4 Study Protocol and Reproducibility*

Each study setting is defined by a dataset, data-generation method, privacy budget, and random seed. We evaluate 96 settings and report averages across completed runs, with confidence intervals where available. PEARL distinguishes successful execution from approval: a completed run that fails an applicable check is a rejected synthetic dataset, not a missing result.

## 4. RESULTS

### *4.1 RQ1: Validity Checks Confirm the Membership Inference Evaluation Is Reliable*

The validity check examines whether the MIA procedure produces misleading above-chance results when no true membership information is present. All four datasets pass this check (Table 1). Randomized-label AUC values are 0.491 for OULAD, 0.505 for UCI 697, and 0.476 for UCI 856 (row-level controls). ACT-MOOC uses a learner-level control; its mean randomized-label AUC across 10 seeds is 0.500. All four values satisfy |AUC – 0.5| < 0.10. Above-chance AUC values in the privacy check are therefore unlikely to result from a flaw in the membership inference evaluation procedure.

### *4.2 RQ2: PEARL Certifies 12 Releases and Documents Where Rejected Releases Fail*

Across 96 study settings, 12 produced synthetic datasets that passed all applicable PEARL checks (Table 2). Nine were tabular datasets from OULAD or UCI 697 using MST or DP-CTGAN, and three were ACT-MOOC datasets using DP-Markov. No UCI 856 setting passed all checks: with only 101 records, DP generation often removed rare outcome groups and left insufficient information for downstream prediction.

Table 1. *Randomized-label validity check for membership-inference evaluation.*

| Dataset | Control Level | Randomized AUC | \|AUC – 0.5\| | Pass |
|---|---|---|---|---|
| OULAD | Row | 0.491 | 0.009 | Yes |
| UCI 697 | Row | 0.505 | 0.005 | Yes |
| UCI 856 | Row | 0.476 | 0.024 | Yes |
| ACT-MOOC | Learner sequence | 0.500 | 0.000 | Yes |
| ASSISTments | Learner | N/A | N/A | Extension only |
| EdNet KT1 | Learner | N/A | N/A | Extension only |

Note. AUC = 0.5 indicates chance-level discrimination. Validity requires |AUC – 0.5| < 0.10. ASSISTments and EdNet KT1 are knowledge-tracing extensions and are not formally validity-certified.

Table 2. *Synthetic datasets approved by PEARL across 96 study settings*

| Dataset | Data type | Synthesizer | ε | MIA AUC | Retention | Final Pass |
|---|---|---|---|---|---|---|
| ACT-MOOC | Sequential | DP-Markov | 1.0 | 0.491 | 0.922 | Yes |
| | | | 5.0 | 0.488 | 0.961 | Yes |
| | | | 10.0 | 0.485 | 0.997 | Yes |
| OULAD | Tabular | DP-CTGAN | 1.0 | 0.462 | N/A (positive) | Yes |
| | | | 5.0 | 0.480 | N/A (positive) | Yes |
| | | | 10.0 | 0.484 | N/A (positive) | Yes |
| | | MST | 1.0 | 0.494 | N/A (positive) | Yes |
| | | | 5.0 | 0.487 | N/A (positive) | Yes |
| | | | 10.0 | 0.490 | N/A (positive) | Yes |
| UCI 697 | Tabular | MST | 1.0 | 0.489 | N/A (positive) | Yes |
| | | | 5.0 | 0.470 | N/A (positive) | Yes |
| | | | 10.0 | 0.474 | N/A (positive) | Yes |

Note. Retention = synthetic-trained macro-F1 / real-trained macro-F1. For OULAD and UCI 697, N/A (positive) denotes passing utility with positive mean ensemble retention. Approved releases pass applicable validity, privacy, utility, and task-specific checks

For the three approved ACT-MOOC datasets, retention values of 0.922, 0.961, and 0.997 indicate that models trained on synthetic data retained 92.2%, 96.1%, and 99.7% of real-data predictive performance at privacy budgets of 1.0, 5.0, and 10.0, respectively. DP-Markov preserved most information required for dropout prediction across all three budgets.

The remaining 84 settings were rejected by at least one PEARL check (Table 3). Forty-five failed privacy because MIA AUC exceeded 0.55. Thirty-four failed predictive usefulness because important outcome groups were missing or performance was near zero, particularly for histogram- and marginal-based methods on UCI 856. Twelve failed the task-specific check because synthetic sequences did not preserve ordered skill–response patterns required by Deep Knowledge Tracing and Self-Attentive Knowledge Tracing. Failure counts are not mutually exclusive because one dataset can fail more than one check.

**Additional Validation with Stronger Membership Inference Attacks.** We examined whether privacy decisions remained stable under Shadow-MIA (an eight-model random forest ensemble) and a likelihood-ratio attack proxy on the OULAD settings in this validation across five seeds. Neither method changed any approval decision (zero of four evaluated settings). The highest Shadow-MIA AUC for an individual seed was 0.585 for PATE-GAN at ε = 10.0, but the mean across seeds was 0.510 (SD = 0.047), so the setting still passed privacy. Within these OULAD settings, stronger attacks did not identify additional privacy failures missed by the initial MIA.

Table 3. *Main failure patterns among rejected configurations.*

| Failure pattern | Count | Main source |
|---|---|---|
| Privacy failure (MIA AUC > 0.55) | 45 | Low-ε configurations across synthesizers |
| Zero or low utility | 34 | Histogram and marginal methods; UCI 856 low-data setting |
| Knowledge-tracing structure loss | 12 | ASSISTments and EdNet KT1 under DP-Markov or DP-HistSeq |
| Label-artifact behavior | 3 | ACT-MOOC DP-HistSeq with inflated retention |

Note. Counts are PEARL diagnostic reasons and are not mutually exclusive; a setting may fail more than one check.

*4.3 RQ3: Task-Specific Checks Separate Sequential Utility from Classifier Utility*

Passing privacy and standard predictive-usefulness checks does not mean that a synthetic dataset supports its intended sequential learning task. For ACT-MOOC, DP-Markov preserved both standard predictive performance and temporal information for sequence modeling across all tested privacy budgets (Table 4). Classifier retention ranged from 0.922 to 0.997, and GRUSeq retention from 0.943 to 0.978.

DP-HistSeq produced the opposite result. Its classifier and GRUSeq retention exceeded 1.0 across all tested budgets, which could appear to outperform real data. Because DP-HistSeq removes event order by design, these scores were treated as label-related artifacts rather than evidence of stronger sequential usefulness.

Table 4. *Task-specific sequential evaluation of ACT-MOOC synthetic data.*

| Synthesizer | ε | Classifier retention | GRUSeq retention | PEARL interpretation |
|---|---|---|---|---|
| DP-Markov | 1.0 | 0.922 | 0.943 | Approved |
| | 5.0 | 0.961 | 0.956 | Approved |
| | 10.0 | 0.997 | 0.978 | Approved |
| DP-HistSeq | 1.0 | 1.100 | 1.302 | Rejected: Artifact |
| | 5.0 | 1.100 | 1.302 | Rejected: Artifact |
| | 10.0 | 1.100 | 1.302 | Rejected: Artifact |

Note. DP-HistSeq is an order-free comparison method; retention > 1.0 is flagged as an artifact. GRUSeq = gated-recurrent sequence model.

The knowledge-tracing evaluation provides a clearer negative example and is treated as a task-specific extension because ASSISTments and EdNet KT1 do not undergo all PEARL checks. As shown in Table 5, models trained on real sequences achieved macro-F1 scores of approximately 0.61, whereas DKT and SAKT learned no meaningful next-response patterns from any tested synthetic dataset across both datasets, both generation methods, all three privacy budgets, and five seeds. DP-HistSeq fails because it removes skill-practice order; DP-Markov preserves skill-identifier order but not the paired skill–correctness information required for next-response prediction. A synthetic dataset may therefore support standard classification while remaining unsuitable for knowledge tracing.

Table 5. *Task-specific evaluation of synthetic knowledge-tracing data.*

| Dataset | Synthesizer | ε | DKT real F1 | DKT synth F1 | SAKT real F1 | SAKT synth F1 | Interpretation |
|---|---|---|---|---|---|---|---|
| ASSISTments | DP-HistSeq | 1, 5, 10 | 0.615 | 0.000 | 0.609 | 0.000 | Order removed |
| | DP-Markov | 1, 5, 10 | 0.615 | 0.000 | 0.609 | 0.000 | Transitions insufficient |
| EdNet KT1 | DP-HistSeq | 1, 5, 10 | 0.617 | 0.000 | 0.615 | 0.000 | Order removed |
| | DP-Markov | 1, 5, 10 | 0.617 | 0.000 | 0.615 | 0.000 | Transitions insufficient |

Note. Synthetic F1 values are exact zeros across five seeds and all tested budgets, indicating that DKT and SAKT cannot learn from these synthetic sequences under this extension. DKT = Deep Knowledge Tracing; SAKT = Self-Attentive Knowledge Tracing.

To check for implementation errors, we evaluated the knowledge-tracing pipeline under five training conditions. Models trained on real sequences achieved macro-F1 ≈ 0.61, and bootstrap samples of real sequences achieved 0.613 (SD = 0.008), confirming a functioning pipeline. Both private and non-private Markov generation produced no learnable next-response patterns because sequences contained skill identifiers without corresponding correctness information. The failure therefore reflects a limitation of the tested generation methods rather than an error in the knowledge-tracing implementation.

The subgroup-fairness analysis examines whether synthetic data preserve similar at-risk predictions across groups (e.g., gender and disability). For OULAD, PEARL rejects a release when the gap between highest and lowest withdrawn-student recall across groups exceeds 0.20. MST with ε = 10.0 passed across all five seeds (mean worst-group recall gap = 0.182, SD = 0.015). PATE-GAN at the same budget passed privacy and predictive usefulness but failed fairness in three of five seeds (mean gap = 0.244, SD = 0.159), with substantial distortions in disability and socioeconomic distributions. The approved MST release also reduced the socioeconomic-group recall gap from 0.261 in the real data to 0.182. Overall privacy and predictive performance can therefore conceal unequal at-risk prediction across student groups.

## 5. Discussion

This study introduced PEARL, a task-aware framework for deciding whether differentially private synthetic educational data can support their intended uses while protecting learners. Privacy protection alone does not establish suitability for educational research or system development. Across 96 settings, only 12 releases passed all applicable PEARL checks. The significance of this finding lies not only in the low approval rate but also in PEARL's ability to explain each approval or rejection, including privacy risks, missing outcome groups, low predictive usefulness, and loss of learning-sequence information.

**Task-Aware Decisions Depend on Dataset Characteristics.** PEARL provides a practical basis for deciding whether a synthetic dataset is suitable for a specific educational purpose by requiring all applicable privacy, predictive-usefulness, and task-specific checks to pass. This is important because successful private data generation depends on the size and structure of the original dataset, the prevalence of important outcome groups, and the selected generation method. In UCI 856, rare outcome groups were frequently lost because the dataset contained only 101 records, while the differing performance of MST and DP-CTGAN across OULAD and UCI 697 showed that no single method is suitable for all educational datasets. PEARL therefore can help identify datasets that appear acceptable overall but fail to preserve at-risk groups or task-specific learning patterns. The randomized-label, shadow-model, and likelihood-ratio analyses further support the privacy decisions, although this evidence remains limited to the evaluated study settings.

**Predictive usefulness does not guarantee sequential suitability.** The ACT-MOOC results show that synthetic data can support sequential learning tasks when the generation method preserves the transitions required by the intended model, as demonstrated by DP-Markov. In contrast, DP-HistSeq achieved high retention despite removing event order, indicating that strong predictive scores may reflect simplified labels or other artifacts rather than meaningful learning patterns. The knowledge-tracing results reinforce this distinction because neither DP-HistSeq nor DP-Markov preserved the paired skill and correctness information required by DKTand SAKT. These findings highlight the need for task-specific evaluation and for future knowledge-tracing synthesizers to preserve both event order and the relationships among skills, responses, and subsequent performance.

**Fairness requires more than overall predictive performance.** Evaluation should consider both subgroup performance gaps and absolute performance within each group; a small gap may be misleading when performance is uniformly poor. PEARL could be extended by combining a maximum subgroup-gap threshold with a minimum subgroup-performance requirement, especially for early-warning systems where errors affect access to support.

**Implications for educational data use.** Approval of synthetic data should be tied to a specified educational task because different applications depend on different data properties. The intended use, applicable privacy and fairness criteria, and task-specific performance requirements should be defined before evaluation. Approval should also apply only to the particular synthetic dataset evaluated, rather than extending to all datasets produced by the same method and privacy budget. PEARL's release ledger supports this process by documenting the generation conditions, evaluation outcomes, and reasons for approval or rejection, thereby reducing the risk that synthetic data are reused for unsupported purposes.

**Limitations and future research.** The evaluation covers six public datasets, a limited set of generation methods, and three broad applications; findings may not generalize to institutional data with different sizes, variables, class distributions, or missingness. PEARL thresholds are study-specific decision rules rather than universal standards and should be adapted to institutional risk and learner consequences. The primary privacy check uses nearest-neighbor MIA; although additional attacks did not change the evaluated OULAD decisions, stronger attacks were not applied to every setting. Future work should broaden datasets, threshold calibration, and adversarial evaluations for high-stakes sharing.

## 6. Conclusion

The findings show that differentially private synthetic educational data should be evaluated according to the task they are expected to support. Privacy protection is necessary, but it does not ensure that important student groups, predictive relationships, subgroup performance, or learning-sequence structures have been preserved. By integrating these considerations into a transparent approval process, PEARL provides a foundation for deciding when synthetic learner data can be used and when further data generation or evaluation is required.

**Acknowledgements**

This work was supported by the University Research Committee Grant (Grant No. 2401102970) at The University of Hong Kong. The opinions, findings, and conclusions expressed in this paper are solely those of the authors.